\documentclass[runningheads]{llncs}

\usepackage{tabularray}
\usepackage{graphicx}
\usepackage[table,xcdraw]{xcolor}
\usepackage{hyperref}
\usepackage{siunitx}
\usepackage{listings}

\lstdefinestyle{mystyle}{
    backgroundcolor=\color{white!20},
    commentstyle=\color{green!50!black},
    keywordstyle=\color{blue},
    numberstyle=\tiny\color{gray},
    stringstyle=\color{orange},
    basicstyle=\ttfamily\footnotesize,
    breakatwhitespace=false,
    breaklines=true,
    captionpos=b,
    keepspaces=true,
    numbers=none,
    numbersep=5pt,
    showspaces=false,
    showstringspaces=false,
    showtabs=false,
    tabsize=2,
}

\usepackage{afterpage}
\usepackage{algorithmic}
\usepackage{array} 
\usepackage{amsmath,amssymb,amsfonts}
\usepackage{booktabs} 
\usepackage{color}
\usepackage{graphbox}
\usepackage{mathtools}
\usepackage{multirow}
\usepackage{textcomp}
\usepackage{tikz}
\usepackage{pifont} 
\usepackage{colortbl}
\usepackage{flushend}
\usepackage[normalem]{ulem}
\useunder{\uline}{\ul}{}
\usepackage{hyperref}
\usepackage[capitalise]{cleveref} 
\usepackage{atbegshi} 
\usepackage{todonotes}
\usepackage{mfirstuc}
\usepackage{csquotes}

\usepackage{textcomp}
\def\review{1} 
\if\review1
\newcommand{\hide}[1]{}
\newcommand{\anonymize}[1]{\textit{redacted for review}}
\else
\newcommand{\hide}[1]{#1}
\newcommand{\anonymize}[1]{#1}
\fi

\begin{document}

\title{Artifact for Service-Level Energy Modeling and Experimentation for Cloud-Native Microservices}
\titlerunning{Artifact for Service-Level Energy Modeling and Experimentation}
\author{Julian Legler
}

\authorrunning{Julian Legler}

\institute{Technische Universität Berlin, Information Systems Engineering \\
\url{https://tu.berlin/ise} \\
\email{julian.legler@tu-berlin.de}}
\maketitle              

\begin{abstract}
Recent advancements enable fine-grained energy measurements in cloud-native environments (e.g., at container or process level) beyond traditional coarse-grained scopes. 
However, service-level energy measurement for microservice-based applications remains underexplored. Such measurements must include compute, network, and storage energy to avoid underestimating consumption in distributed setups.
We present GOXN (Green Observability eXperiment eNginE), an energy experimentation engine for Kubernetes-based microservices that quantifies compute, network, and storage energy at the service level. 
Using GOXN, we evaluated the OpenTelemetry Demo under varying configurations (monitoring, tracing, service mesh) and steady synthetic load, collecting metrics from Kepler and cAdvisor.
Our additive energy model derives service-level energy from container-level data. Results show that excluding network and storage can underestimate auxiliary-service energy by up to \SI{63}{\percent}, and that high tracing loads shift energy dominance toward network and storage.

\end{abstract}
%


\section{Introduction}\label{ch:intro}
GOXN \cite{legler_goxn_2025} is an open-source experimentation engine that automates the execution of energy experiments for cloud-native microservice applications deployed on Kubernetes (K8s). It provisions an observability stack, deploys the System Under Evaluation (SUE), applies treatments, generates load, collects metrics, and computes per-service energy consumption considering compute, network, and storage. GOXN enables researchers and practitioners to conduct reproducible energy experiments in cloud-native environments. 
It is designed to be extensible, allowing users to adapt it to their specific needs and experiment setups.

GOXN is an extension of OXN (Observability eXperiment eNginE)~\cite{borges_informed_observability_decisions_2024}. 
OXN was previously used to help practitioners make informed observability decisions by quantifying the impact of different observability configurations on failure detection metrics like Time To Detect (TTD). 
GOXN significantly extends OXN by adding energy measurement capabilities, specifically focusing on service-level energy consumption in microservice architectures. 
For this, the core architecture of OXN was completely refactored, to elevate OXN from a locally run Docker-deployments only to a K8s-native tool.
Additionally, it introduces energy measurement capabilities that allow for fine-grained tracking of energy consumption across different services and components within a microservice architecture deployed on K8s. 
For this purpose, GOXN integrates with Kepler~\cite{amaral_kepler_2023}, a K8s-native energy measurement tool that provides per-container energy consumption data.

The utilized energy model approach is a component level additive model as described by \cite{Dayarathna_energy_consumption_modeling_2016}. 
It computes service-level energy from container-level metrics based on compute energy from Kepler and network/storage energy derived from proxy metrics using energy-intensity factors \cite{aslan_kwh_per_gb_2018,pijnacker_container_level_observability_2025,al_kez_exploring_2022}. 
This allows for a comprehensive view of energy consumption, capturing the contributions of compute, network, and storage to the overall energy footprint of each service.

This document describes the software, data, and procedures needed to independently reproduce and verify the results of our study on service-level energy modeling for cloud-native microservices. 
The artifact has two aims: (A) enable end-to-end re-execution of the experiments on a Kubernetes (K8s) cluster, and (B) reproduce the reported tables and figures from preserved raw data alone. We provide public links, concrete steps and explicit software/hardware assumptions.


\section{Artifact Contents}
\textbf{A: Executable Framework (GOXN)}\footnote{Available at \url{https://github.com/JulianLegler/goxn}} GOXN (Green Observability eXperiment eNginE) automates experiment execution: provisioning an observability stack, deploying the SUE, applying treatments, generating load, collecting metrics, and computing per-service energy.

\smallskip
\noindent\textbf{B: Replication Package (Data \& Notebooks)}\footnote{Available at \url{https://github.com/JulianLegler/goxn-replication-package}} Contains the exact configuration files used, raw experiment outputs (E1--E3), and Jupyter notebooks that regenerate all tables and figures.


\section{Reproduction Path A: End-to-End Re-Execution on Kubernetes}
The objective of reproduction path A is to deploy a full experiment setup including a freshly set up Kubernetes Cluster, the SUE and observability stack, deploy GOXN, run all experiments conducted in the paper and compute the service-level energy.
Due to the needed low level access to hardware information, the use of an VM or cloud provider is not recommended for this path and might not work at all. 
This limitation is mainly due to the fact that Kepler needs access to hardware information that is often not available in VMs or cloud providers.
If you want to avoid setting up a K8s cluster on a bare metal machine (or locally if you have the necessary prerequisites), please refer to reproduction path B (Data-Only Reproduction).

\subsubsection{Prerequisites} Linux OS (e.g. Debian), \texttt{kubectl}, \texttt{helm}; Python~\(\geq\)~3.10; Jupyter; cloned the goxn github repository via: \lstinline{git clone https://github.com/Julian Legler/goxn}

\subsubsection{Configure the Kubernetes Cluster}
Using MicroK8s \footnote{Available at \url{https://microk8s.io/}} to create a Kubernetes Cluster. Install the default StorageClass (OpenEBS example below) to enable PersistentVolumeClaims (PVCs). For this, run \lstinline{sh 1_setup_storage_provider.sh}.

Deploy Kepler Helm chart once in the cluster. Kepler runs as a DaemonSet and collects per-container energy consumption data and exposes it via Prome\-theus. This can be done by running \lstinline{sh 2_setup_kepler.sh}

\subsubsection{Install GOXN and ensure dependencies are installed}
Clone the GOXN repository and install it in a Python virtual environment via \lstinline{sh 3_setup_goxn.sh}

\subsubsection{Run experiments}
We provide a script that runs all experiments sequentially (\textit{run\_experiment\_suite.sh}). It uses the configuration files from the replication package (B) to deploy the exact same scenarios as in the paper.
This script performs the following steps for each scenario:
\begin{itemize}
  \item Ensure a clean state by uninstalling previous deployments.
  \item Set up the observability stack (kube-prometheus-stack).
  \item Setup the SUE (OpenTelemetry Demo).
  \item Run GOXN to apply treatments, generate load, collect metrics.
  \item Persist raw results.
  \item Start the next scenario.
\end{itemize}

\subsubsection{Raw outputs pre-processing}
GOXN persists raw outputs in a HDF5 file in the root directory named \texttt{store.h5}. It contains all raw metrics collected from Prometheus/cAdvisor and Kepler. 
Additionally, GOXN generates a report as a YAML file for each experiment in the \texttt{reports/} directory. 
This report contains some metadata about the experiment run (e.g., start/end time, treatment applied) and is crucial for the further processing of the raw metrics.
Storage metrics are separately stored in the directory \texttt{storage\_snapshots/} as CSV files.

The next step is a combination of multiple pre-processing steps and the actual transformations of the proposed energy model. 
This step results are then in a format that can be consumed by the notebooks in the replication package (B) to generate all the figures and data from the paper.

This step is automated by the Jupyter Notebook \textit{data\_processing.ipynb}. 
It generates the following important \texttt{CSV} files in the root directory:
\begin{itemize}
  \item \texttt{cadvisor\_storage\_usage\_writes\_all\_absolute\_bytes.csv} - Used later in the energy model as $B_{\mathbf{storage}}$
  \item \texttt{cadvisor\_network\_bytes\_received\_all\_absolute\_bytes.csv} - Used later in the energy model as $B_{\mathbf{network}}$
  \item \texttt{pods\_kepler\_joules\_all\_absolute\_joules.csv} - Used later in the energy model as $E_{\mathbf{compute}}$
  \item \texttt{energy\_totals\_*.csv} - Contains the computed service-level energy for each scenario and service using the energy model described in the paper.
\end{itemize}

These files can then be used as input to the notebooks in the replication package (B) to regenerate all tables and figures from the paper.


\section{Reproduction Path B: Data-Only Reproduction (No Cluster)}
The objective of this path is to regenerate all tables/plots from the preserved results delivered within the goxn replication package.

\subsubsection{Prerequisites} Python~\(\geq\)~3.10

\subsubsection{Retrieve the replication package}
All needed files are provided in the replication package (B). 
It contains the exact configuration files used, raw experiment outputs (E1--E3), and Jupyter notebooks that regenerate all tables and figures from the experiment outputs. Retrieve it via \lstinline{git clone https://github.com/JulianLegler/goxn-replication-package}

\subsubsection{Generate tables and figures from raw results}
The experiment data is stored in the directory \texttt{experiment\_data/} as CSV, for human readability, and in their raw unprocessed form as HDF5 files.
The notebooks in the directory \texttt{evaluation/} consume these files directly and regenerate all tables and figures from the paper.
The evaluation notebooks are organized as follows:
\begin{itemize}
  \item \texttt{energy-evaluation.ipynb} - Regenerates all tables and figures related to service-level energy from the paper.
  \item \texttt{confidence\_interval.ipynb} - Can be used for further input and understanding of the data by computing 95 \% confidence intervals and means for the measured metrics.
\end{itemize}

\subsubsection{Scenario-to-Configuration Mapping}
The replication package (\texttt{configu\-rations/}) contains the exact YAMLs used in the paper. Table \ref{tab:config-mapping} can be used to identify which configuration file corresponds to which scenario in the paper.
\begin{table}[]
\caption{Mapping of paper scenarios to configuration files in replication package.}
\label{tab:config-mapping}
\centering
\begin{tblr}{
  colspec = {l l},
  row{1} = {font=\bfseries},
  hlines, vlines
}
Paper Scenario & Configuration file name \\
Baseline & \texttt{recommendation\_k8\_base\_1m\_otel\_persistence.yaml} \\
Monitoring Medium & \texttt{recommendation\_k8\_base\_1m\_otel\_persistence\_scrape\_30s.yaml} \\
Monitoring High & \texttt{recommendation\_k8\_base\_1m\_otel\_persistence\_scrape\_5s.yaml} \\
Tracing Low & \texttt{recommendation\_k8\_base\_5\_percent\_persistence.yaml} \\
Tracing Medium & \texttt{recommendation\_k8\_base\_10\_percent\_persistence.yaml} \\
Tracing High & \texttt{recommendation\_k8\_base\_50\_percent\_persistence.yaml} \\
Service Mesh & \texttt{recommendation\_k8\_base\_1m\_otel\_persistence\_istio.yaml} \\
\end{tblr}
\end{table}

\subsubsection{Add new data from end-to-end execution}
If you have executed the end-to-end path (A) and have the processed CSV files from the \texttt{data\_processing.ipynb} notebook, you can create a new directory in \texttt{experiment\_data/} (e.g. \texttt{E4}) and copy the CSV files there.
Make sure to name and structure the files identically to the existing directories (E1--E3).
In the \texttt{energy-evaluation.ipynb} notebook, change the \texttt{experiment} variable in the second code cell to point to your new directory (e.g. \texttt{E4}).
Then, re-run all cells to regenerate the tables and figures from your new data.


\section{Extendability and Configurability}
GOXN is designed to be extensible. Users can adapt it to their specific needs by adding new SUEs or Treatments. 
\begin{itemize}
  \item \textbf{Add new Treatments.} Treatments are defined in \texttt{treatments.py}. Newly added treatments have to be registered in the \texttt{runner.py} file in the \texttt{treatment \_keys} list. Follow the structure of existing Treatments for guidance.
  \item \textbf{Custom Load Generation.} Modify or extend the load generation scripts in the SUE. GOXN uses locust for load generation. The locust files are located in the \texttt{locust/} folder and are easily modifiable.
  \item \textbf{Add new SUEs.} Alter the \textit{run\_experiment\_suite.sh} script to include the proper setup and cleanup process for the new SUE. Change the load generation files in \textit{locust/}. Change the experimentation configuration to reflect the new SUE as the \texttt{loadgen.target}. Follow the structure of existing SUEs for guidance.
  \item \textbf{Custom Metrics Collection.} Metrics for collection are defined in the experiment configuration files and can easily be extended using PromQL. Modify the \texttt{responses} section to add or remove metrics as needed. Follow the structure of existing metrics for guidance.
\end{itemize}


\section{Data Origins and Provenance}
All raw measurements were produced by GOXN during the described experiments. Metrics are collected from Prometheus/cAdvisor and Kepler at container level and aggregated to service level using the model above.


\section{Experiment Environment}
The experiments were conducted on a dedicated server with the following specifications:
\begin{itemize}
  \item \textbf{Hardware:} Intel(R) Xeon(R) E-2176G CPU @ 3.70GHz, 64 GB DDR4.
  \item \textbf{Operating System:} Debian 6.1.119-1.
  \item \textbf{Kubernetes:} MicroK8s v1.32.3 revision 8148.
  \item \textbf{Helm:} v3.18.2.
  \item \textbf{Kepler:} Helm chart version 0.6.0, Kepler release-0.8.0.
  \item \textbf{SUE:} OpenTelemetry Demo Helm chart version 0.36.4.
  \item \textbf{Observability Stack:} kube-prometheus-stack Helm Chart version 62.5.1
\end{itemize}


\section{License}
GOXN and the replication package are released under the GPLv3 License. Third-party dependencies may have different licenses.



\bibliographystyle{src/splncs04}
\bibliography{refs}

\end{document}